\documentclass[conference]{IEEEtran}
\IEEEoverridecommandlockouts
\usepackage{cite}
\usepackage{
amsmath,amssymb,amsfonts}
\usepackage{algorithmic}
\usepackage{graphicx}
\usepackage[inkscapelatex=false]{svg}
\usepackage{tabularx}
\usepackage{textcomp}
\usepackage{xcolor}
\usepackage{multirow} 
\usepackage{array}  
\usepackage{tikz}
\usepackage{float} 
\usepackage{hyperref}
\usepackage{xcolor}
\usepackage{orcidlink}
\hypersetup{
    colorlinks=true,
    linkcolor=blue,
    citecolor=blue,
    filecolor=magenta,      
    urlcolor=cyan,
    pdftitle={},
    pdfpagemode=FullScreen,
}

\usetikzlibrary{shapes.geometric, arrows, positioning}
\def\BibTeX{{\rm B\kern-.05em{\sc i\kern-.025em b}\kern-.08em
    T\kern-.1667em\lower.7ex\hbox{E}\kern-.125emX}}
\begin{document}
\graphicspath{{1_Conference_Papers_x4/1b_PosiPap_INDIN25Q3_WS01/figures/}}


\title{CFTel: A Practical Architecture for Robust and Scalable Telerobotics with Cloud-Fog Automation}

\author{\IEEEauthorblockN{
    Thien Tran\IEEEauthorrefmark{1},    
    Jonathan Kua\IEEEauthorrefmark{1},  
    Minh Tran\IEEEauthorrefmark{2}\IEEEauthorrefmark{5}, 
    Honghao Lyu\IEEEauthorrefmark{3},   
    Thuong Hoang\IEEEauthorrefmark{1}  
    and Jiong Jin\IEEEauthorrefmark{4}  
    } \\
\IEEEauthorblockA{
    \IEEEauthorrefmark{1}Deakin University, Australia; 
    \IEEEauthorrefmark{2}RMIT University, Vietnam; 
    \IEEEauthorrefmark{3}Zhejiang University, China; \\ 
    \IEEEauthorrefmark{4}Swinburne University of Technology, Australia; 
    \IEEEauthorrefmark{5}University of Tasmania, Australia \\
    {\{peter.tran, jonathan.kua, thuong.hoang\}@deakin.edu.au};\\
    {minh.tranquang@rmit.edu.vn};
    {lvhonghao@zju.edu.cn};
    {jiongjin@swin.edu.au} \\
    }\\
}

\maketitle

\begin{abstract}
Telerobotics is a key foundation in autonomous Industrial Cyber-Physical Systems (ICPS), enabling remote operations across various domains. However, conventional cloud-based telerobotics suffers from latency, reliability, scalability, and resilience issues, hindering real-time performance in critical applications. Cloud-Fog Telerobotics (CFTel) builds on the Cloud-Fog Automation (CFA) paradigm to address these limitations by leveraging a distributed Cloud-Edge-Robotics computing architecture, enabling deterministic connectivity, deterministic connected intelligence, and deterministic networked computing. This paper synthesizes recent advancements in CFTel, aiming to highlight its role in facilitating scalable, low-latency, autonomous, and AI-driven telerobotics. We analyze architectural frameworks and technologies that enable them, including 5G Ultra-Reliable Low-Latency Communication, Edge Intelligence, Embodied AI, and Digital Twins. The study demonstrates that CFTel has the potential to enhance real-time control, scalability, and autonomy while supporting service-oriented solutions. We also discuss practical challenges, including latency constraints, cybersecurity risks, interoperability issues, and standardization efforts. This work serves as a foundational reference for researchers, stakeholders, and industry practitioners in future telerobotics research.
\end{abstract}


\section{Introduction}\label{sec:sec1}
Telerobotics is a critical application of next-generation Industrial Cyber-Physical Systems (ICPS), enabling remote control of robotic systems in complex environments within Industry 4.0~\cite{Jin2025_04908, Adil2025_3486659, Ji2025_102927, Jin2024_3272696, Lopez2022_DT}. Conventional cloud-based telerobotics, which relies on centralized processing, faces significant latency and network congestion, limiting real-time performance in critical applications of industrial automation~\cite{Afrin2021_3061435, Lopez2022_DT}. Cloud-Fog Automation (CFA) is a new paradigm that addresses these challenges by integrating a distributed Cloud-Edge-Robotics architecture with deterministic networking. It incorporates Edge Intelligence (EI), Embodied AI (EAI), and Digital Twins (DTs) to deliver scalable, low-latency, autonomous, and intelligent telerobotic systems~\cite{Hu2024_10595837, Himeur2024_101035}. This framework leverages the combined capabilities of Cloud-Fog computing, which emphasizes deterministic communication, computing, and control, alongside scalable, service-oriented solutions~\cite{Afrin2021_3061435}.

Leveraging the dynamic and network-centric features of CFA, Cloud-Fog Telerobotics (CFTel) enhances conventional telerobotics by distributing computational workloads across cloud, edge, and robotics layers~\cite{Jin2024_3272696, Khan2020_1010}. The cloud layer supports big data analytics, AI model deployment, and multi-robot coordination. While edge nodes enable low-latency AI inference, time-sensitive operations, and decentralized allocation of computing load using fog computing, the robotics layer, powered by EAI, ensures autonomous adaptation to dynamic environments~\cite{Lyu2024_3407051, Hu2024_10595837}. Deterministic networking technologies, such as 5G Ultra-Reliable Low-Latency Communication (URLLC) and Time-Sensitive Networking (TSN), underpin real-time control, ensuring high-reliability networks~\cite{Luo2023_3228558, Ma2019_2907245}. These advancements facilitate applications in smart manufacturing, agriculture, logistics, automated warehousing, healthcare, and disaster management, supporting service-oriented models that enhance scalability and flexibility in Industry 4.0.

Despite technological advancements, CFTel faces challenges, including latency constraints, cybersecurity risks, interoperability issues, and scalability limitations, necessitating standardized architectures and robust AI-driven frameworks \cite{Xiao2024_111974}. This paper synthesizes recent advancements to empower CFTel, highlighting its role in enabling scalable, low-latency, autonomous, and AI-driven telerobotics. In summary, we make the following contributions:

\begin{itemize}
    \item Review state-of-the-art enabling technologies for telerobotics, focusing on EI, EAI, and DTs.
    \item Propose a novel Cloud-Fog Telerobotics architecture based on the Cloud-Fog Automation.
    \item Outline challenges and future research directions for AI-driven control systems to enhance robust telerobotics.
\end{itemize}

This paper is structured as follows: Section~\ref{sec:sec2} presents the CFTel architecture, Section~\ref{sec:sec3} details technologies, Section~\ref{sec:sec4} describes industrial applications, Section~\ref{sec:sec5} addresses practical challenges, and Section~\ref{sec:sec6} concludes with future directions.

\begin{figure*}[!ht]
    \centering
    \includegraphics[width=0.84\textwidth]{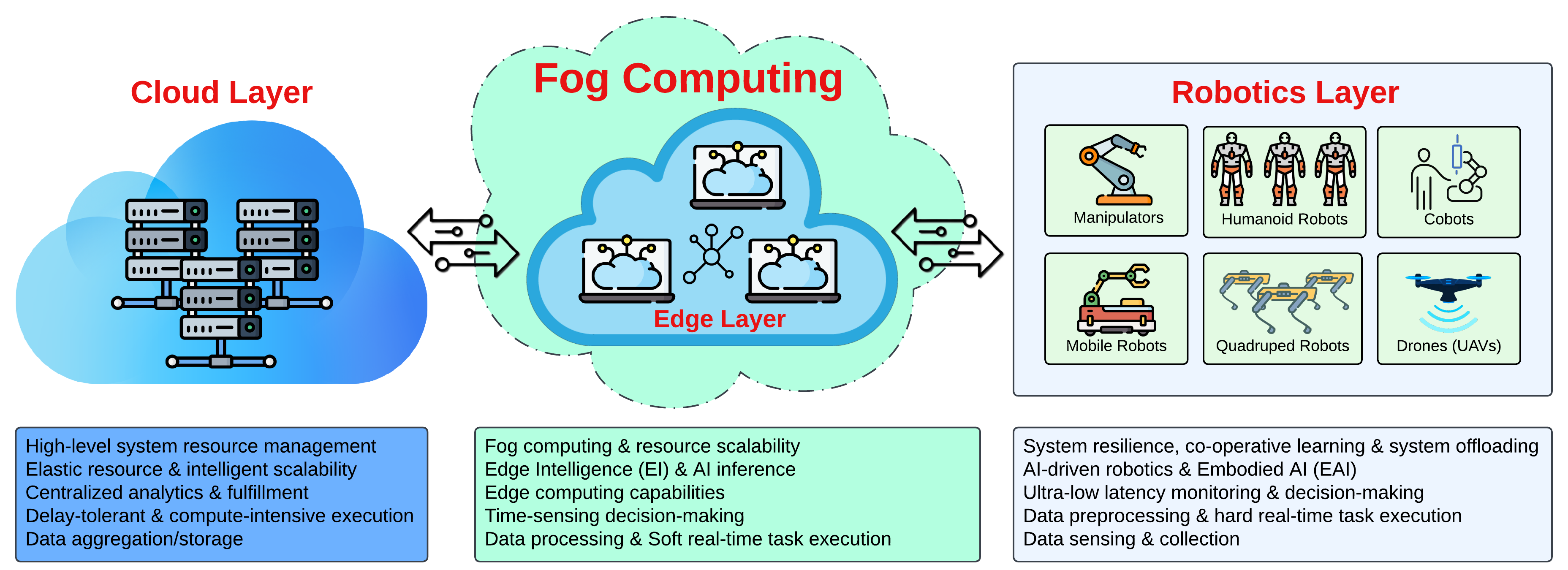}
    \caption{The distributed computing architecture leveraging Cloud-Fog Automation for Cloud-Fog Telerobotics.}
    \label{fig1}
\end{figure*}

\section{Cloud-Fog Telerobotics Architecture}\label{sec:sec2}
CFTel employs a hierarchical Cloud-Edge-Robotics architecture to distribute computational and operational workloads, ensuring deterministic communication, computing, and control~\cite{Jin2024_3272696, Yang2024}. Table~\ref{tab1} compares the limitations of conventional cloud-based telerobotics with the proposed architecture by integrating distributed, decentralized, and localized processing. This approach enables computational load balancing, time-sensitive decision-making, security, scalability, and resilience for AI-driven telerobotics. It also enhances real-time performance, resilience, and scalability, and supports service-oriented solutions such as Robots-as-a-Service (RaaS), Control-as-a-Service (CaaS), and Manufacturing-as-a-Service (MaaS)~\cite{Afrin2021_3061435}. Figure~\ref{fig1} illustrates the overall architecture of CFTel, and each layer plays a key role as follows:

\begin{itemize}
    \item \textbf{Cloud Layer:} Centralizes high-level processing for ICPS analytics, AI development and training, and multi-robot coordination, enabling MaaS and RaaS.
    \item \textbf{Edge Layer:} Enables low-latency decision-making, AI inference for time-sensitive operations, supporting CaaS.
    \item \textbf{Robotics Layer:} Provides ultra-low-latency control, resilience, and autonomy with EAI for telerobotics in ICPS.
\end{itemize}

\subsection{Cloud Layer}\label{sec:sec2A}
The cloud layer serves as the centralized hub for high-level processing, providing extensive data storage, big data analytics, and large-scale AI model training. It facilitates global intelligence through tasks such as predictive maintenance, multi-robot task allocation, and fleet-wide optimization. The cloud supports scalable, vendor-independent models such as MaaS, enabling seamless integration of heterogeneous robotic systems~\cite{Afrin2021_3061435, Lyu2024_3407051, Himeur2024_101035}. It also hosts DTs, virtual replicas of physical systems, for real-time simulation and diagnostics~\cite{Tao2019_S209580991830612X}, which are critical for applications such as precision agriculture \cite{Soumik2024_CAS}. However, reliance on remote data centers introduces latency, limiting its suitability for time-critical tasks such as telerobotic surgery or disaster response. The cloud layer interfaces with lower layers to delegate time-sensitive operations, ensuring balanced workload distribution~\cite{Afrin2021_3061435}.

\subsection{Edge Layer}\label{sec:sec2B}
The edge layer enables low-latency decision-making and AI inference by processing data closer to robotic systems. Deployed in industrial environments, edge nodes execute lightweight AI models for time-sensitive tasks, including quality inspection, path planning, system failure detection, and data encryption~\cite{Gong2024_108906, Zhang2024_104131}. This layer incorporates fog computing principles, often referred to as EI, which reduces cloud dependency, enhances security, and optimizes bandwidth~\cite{Himeur2024_101035}. Edge computing supports CaaS, enabling modular control services for commercial applications. By distributing computational loads across multiple edge nodes, this layer enhances network reliability and resilience, particularly in dynamic environments where cloud connectivity may be intermittent~\cite{Xiao2024_111974}. Traditional time-sensitive control functions, typically managed by Programmable Logic Controllers (PLCs), can be migrated to the Edge Layer using EI. EI can execute control logic with AI-driven models, such as reinforcement learning for adaptive control, while leveraging TSN and 5G URLLC to ensure deterministic performance with sub-millisecond response times~\cite{Yang2024}. However, challenges include adapting legacy PLC code to AI-based systems, which requires careful validation to maintain reliability and safety in industrial settings~\cite{Hu2024_10595837}.

\subsection{Robotics Layer}\label{sec:sec2C}
The robotics layer delivers ultra-low-latency execution, utilizing EAI for real-time robotic control and autonomy~\cite{Jin2025_04908}. Comprising physical robotic systems, this layer processes sensor inputs, such as LiDAR, cameras, and haptic feedback, to enable millisecond-level responses~\cite{Gong2024_108906}. EAI empowers robots to perceive, learn, and adapt to dynamic environments. This layer enhances resilience through localized intelligence, minimizing dependency on upper layers~\cite{Jin2024_3272696, Himeur2024_101035}. The integration with edge nodes ensures seamless data flow, supporting scalable autonomous solutions, such as RaaS and CaaS.

\subsection{Integration with Existing Industrial Systems}\label{sec:sec2D}
Integrating CFTel architecture into existing industrial systems is critical for practical deployment in Industry 4.0. Legacy systems, such as PLCs and Supervisory Control and Data Acquisition systems, can be interfaced with CFTel using OPC Unified Architecture (OPC-UA). OPC-UA enables seamless data exchange between legacy systems and CFTel’s Edge and Cloud layers, ensuring compatibility and real-time monitoring~\cite{ladegourdie2022performance,Ji2025_102927}. Additionally, middleware solutions abstract the underlying complexities of heterogeneous systems, providing a unified interface for data aggregation and control. This approach ensures that CFTel can be incrementally~adopted~in~existing industrial setups without requiring~a~complete~overhaul.

\subsection{Performance Evaluation Approach}\label{sec:sec2E}
To assess CFTel’s practical relevance, performance evaluation can focus on key metrics such as latency, throughput, and reliability. Latency, as shown in Table~\ref{tab1}, is reduced to 5-10 ms in CFTel compared to 50-100 ms in conventional cloud-based telerobotics. Throughput can be measured by the number of robotic tasks processed per second, while reliability can be evaluated through the percentage of successful task completions under varying network conditions. A simulation-based approach using tools such as NS-3 or OMNeT++ can model CFTel’s architecture in real-world applications such as multi-robot coordination in logistics or networked robotic systems in smart manufacturing for validating CFTel’s advantages.

\begin{table*}[h]
\renewcommand{\arraystretch}{1.3}
\centering
\caption{Comparison of Conventional Cloud-Based Telerobotics vs. Cloud-Fog Telerobotics}
\label{tab1}
\begin{tabularx}{\textwidth}{|p{1.5cm}|X|X|}
\hline
\textbf{Feature} & \textbf{Conventional Cloud-Based Telerobotics} & \textbf{Cloud-Fog Telerobotics} \\
\hline
Latency & High due to centralized processing and geographic server location & Lower with distributed processing across the Cloud-Fog continuum \\
\hline
Scalability & Limited by centralized infrastructure and network congestion & Enhanced via fog computing, incorporated with cloud computing \\
\hline
Real-Time Control & Delayed, dependent on network stability, suitable for global system monitoring, analytics, and resource management & Fast, cyber-secure, distributed processing, cost-effective, supports time-sensitive tasks close to industrial robotic systems \\
\hline
Security & Vulnerable due to centralized data storage and targeted attacks & Improved with localized encryption and fog-edge intelligence \\
\hline
Resilience & Susceptible to cloud failures, network congestion, and cyber-attacks & More robust with decentralized intelligence, cloud-independent \\
\hline
\end{tabularx}
\end{table*}

\section{Enabling Technologies for CFTel}\label{sec:sec3}
The implementation of CFTel relies on advanced technologies that ensure deterministic communication, real-time computing, and intelligent control. This section examines three critical enabling technologies: Network Infrastructure, EI and EAI, and DTs. These technologies underpin the hierarchical architectural layers, facilitating scalable, low-latency, and autonomous telerobotic operations. By integrating these technologies, CFTel achieves deterministic performance and service-oriented solutions as presented in Table~\ref{tab2}.

\begin{table}[!b]
\centering
\caption{Emerging Technologies and Their Key Roles in CFTel}
\begin{tabularx}{\columnwidth}{|p{1.3cm}|X|p{2.5cm}|p{1cm}|}
\hline
\textbf{Technology} & \textbf{Key Role} & \textbf{Applications} & \textbf{Layer} \\
\hline
5G/URLLC & Provides ultra-low-latency, high-reliability communication & Precise control, wireless communication, advanced telerobotics & Cloud, Edge, Robotics  \\
\hline
Wi-Fi 6/7 & Enhances throughput in dense environments & Autonomous logistics, manufacturing & Cloud, Edge, Robotics \\
\hline
TSN & Ensures deterministic, low-jitter communication & Automation coordination, time-sensitive operations & Edge, Robotics \\
\hline
DTs & Optimizes performance, predicts failures & Manufacturing, resource management & Cloud, Edge \\
\hline
EI & Enables real-time AI inference & Quality control, decision-making & Edge \\
\hline
EAI & Facilitates autonomous robot adaptation & Network connectivity, system resilience & Robotics \\
\hline
\end{tabularx}
\label{tab2}
\end{table}

\subsection{Network Infrastructure}\label{sec:sec3A}
A robust network infrastructure serves as the backbone of CFTel, providing deterministic, low-latency communication. Several advanced technologies contribute to this framework:

\begin{itemize}
    \item \textbf{Time-Sensitive Networking:} TSN extends Ethernet to ensure deterministic data transmission with minimal jitter, which is critical for synchronized operations~\cite{Jin2025_04908, Hu2024_10595837}.
    \item \textbf{5G and Beyond:} 5G deliver ultra-low-latency and high-reliability communication, supporting long-distance telerobotics. Emerging 6G promises further reductions in latency, ideal for time-sensitive applications~\cite{Jin2024_3272696, Lopez2022_DT, Ma2019_2907245}.
    \item \textbf{Wi-Fi 6/7:} These standards offer high throughput, low latency, and improved performance, complementing 5G and TSN by enhancing wireless connectivity, reducing congestion, and supporting real-time data exchange~\cite{Hu2024_10595837}.
\end{itemize}

\subsection{Edge Intelligence and Embodied AI}\label{sec:sec3B}
EI and EAI elements synergize within the robotics layer, leveraging fog-edge computing for low-latency inference and enabling adaptive, resilient telerobotic systems across ICPS applications~\cite{Adil2025_3486659, Lopez2022_DT, Himeur2024_101035}.

\begin{itemize}
    \item \textbf{Edge Intelligence:} EI deploys AI models at edge nodes, processing data locally to minimize latency and reduce cloud dependency. Convolutional Neural Networks can be deployed for time-sensitive tasks such as quality inspection and system predictive maintenance. Federated learning enables decentralized model training across~edge, preserving privacy and improving efficiency~\cite{Zhang2025_3528569, Yang2024_S2452414X2400089X}.
    \item \textbf{Embodied AI:} EAI integrates AI into physical systems, enabling perception, learning, and adaptation~\cite{Luo2023_3228558}. Through sensor fusion and reinforcement learning, EAI facilitates adaptive autonomy, allowing robots to optimize and perform complex tasks with minimal human input while enhancing human-robot interaction~\cite{Tran2025_S2452414X24001869, Nguyen2024_10912837}.
\end{itemize}

\subsection{Digital Twins}\label{sec:sec3C}
DTs are virtual representations of physical robotic systems, mirroring their real-time state and behavior to optimize performance and reliability in CFTel. Hosted primarily in the cloud or edge layers, DTs integrate data from robotics and edge nodes to provide real-time insights~\cite{Tao2019_S209580991830612X, Tran2025_S2452414X24001869}. {Data~management~employs~a~publish-subscribe~model,~such as MQTT,~ensuring~efficient~real-time~updates~with~minimal~overhead~\cite{Luo2023_3228558}.

\begin{itemize}
    \item \textbf{Predictive Maintenance:} DTs analyze historical and real-time data to predict potential failures, enabling proactive maintenance scheduling to minimize downtime.
    \item \textbf{System Optimization:}  DTs facilitate simulation, testing, and optimization of robotic configurations, enabling parameter refinement before real-world deployment.
    \item \textbf{Remote Diagnostics and Monitoring:} DTs allow operators to visualize robot states and detect anomalies remotely via immersive eXtended Reality interfaces.
\end{itemize}

\section{Industrial Applications of CFTel}\label{sec:sec4}
This section explores CFTel's transformative impact on smart manufacturing and agriculture, logistics and warehousing, healthcare, and disaster management, detailing specific use cases, benefits, and challenges as presented in Table~\ref{tab3}. Each subsection highlights the enabling technologies that drive innovation and support service-oriented solutions.

\begin{table*}[h]
\renewcommand{\arraystretch}{1.3}
\centering
\caption{Summary of CFTel Industrial Applications, Enabling Technologies, Benefits, and Challenges}
\label{tab3}
\begin{tabularx}{\textwidth}{|p{2.5cm}|X|X|p{3.5cm}|p{2.3cm}|}
\hline
\textbf{Domain} & \textbf{Key Use Cases} & \textbf{Enabling Technologies} & \textbf{Benefits} & \textbf{Challenges} \\
\hline
Smart Manufacturing & Adaptive automation, robotic coordination, quality inspection & Cloud-Fog computing, EI, EAI, Wi-Fi 6, TSN & Reduced downtime, minimized waste & System integration \\
\hline
Agriculture & Spraying, fertilizing, crop monitoring, livestock monitoring  & Cloud computing, EI, autonomous drones & Optimized yields, real-time decision-marking & Rural connectivity \\
\hline
Logistics & Robotic fleet management, navigation, inventory management  & 5G URLLC, Wi-Fi 6, DTs, EI & Increased throughput, safety & Cybersecurity \\
\hline
Healthcare & Robotic surgery, assistive robots & Cloud-Fog computing, EI, EAI & Precision, personalized care & Data privacy \\
\hline
Disaster Management & Infrastructure inspection, search-and-rescue, swarm coordination & 5G/URLLC, Edge computing, EI, EAI & Responsiveness, life-saving operations & Bandwidth limitations \\
\hline
\end{tabularx}
\end{table*}

\subsection{Smart Manufacturing and Agriculture}\label{sec:sec4A}
The distributed CFTel architecture advances smart manufacturing by enabling adaptive automation, remote monitoring, and predictive maintenance in advanced autonomous ICPS. Cloud-based AI analyzes sensor data to predict equipment failures, reducing factory downtime and executing high-level decision-making~\cite{Jin2024_3272696}. Robotic arms in automotive assembly lines use fog computing for real-time quality inspection, detecting defects instantly to minimize waste. Edge nodes empower robots with adaptive control, allowing immediate responses to production anomalies, thus enhancing operational efficiency and cost-effectiveness. These capabilities ensure high reliability, scalability, and resilience, supporting MaaS, RaaS, and CaaS for vendor-independent deployments~\cite{Jin2025_04908, Zhang2024_104131}.

In agriculture, CFTel facilitates precision farming through autonomous drones and robots powered by EI and EAI~\cite{Gong2024_108906, Sibona2025}. These systems perform tasks such as pest detection, crop monitoring, and soil analysis, processing data locally to optimize yields in real time~\cite{Xiao2024_111974}. Drones equipped with LiDAR and cameras monitor crop health, adapting irrigation based on edge-driven insights, reducing reliance on centralized cloud processing~\cite{Soumik2024_CAS}. CFTel's architecture enhances scalability, enabling farmers to manage large fields efficiently. Challenges include integrating heterogeneous robotic systems and ensuring robust connectivity in rural areas, necessitating standardized protocols and advanced networking solutions~\cite{Lyu2024_3407051}.

\subsection{Logistics and Warehousing}\label{sec:sec4B}
CFTel transforms logistics and warehousing by enabling real-time coordination of autonomous robotic fleets, boosting efficiency and throughput~\cite{Xiang2024_3264289}. The cloud layer optimizes fleet-wide operations using AI-driven algorithms, while edge nodes handle localized tasks. Leveraging 5G URLLC, robots dynamically navigate for sub-millisecond coordination with other robots and workers, enhancing safety and speed~\cite{Hu2024_10595837}. DTs simulate warehouse workflows, predicting bottlenecks and optimizing layouts, further improving operational efficiency~\cite{Tao2019_S209580991830612X}.

CFTel’s distributed architecture supports CaaS, allowing modular control services for scalable logistics operations~\cite{Afrin2021_3061435}. The communication paradigm of CFTel, combined with EI, ensures robots adapt to dynamic environments. {This architecture inherently supports scalability by offloading computational tasks to edge nodes, reducing bottlenecks in high-density robotic networks. Technologies such as 5G URLLC and EI enable efficient management of large-scale fleets, ensuring low-latency communication and decentralized decision-making. However, handling high-density robotic nodes, ensuring cybersecurity and safety, in a logistics hub requires further investigation into network capacity and swarm intelligence~\cite{Xiang2024_3264289}.

\subsection{Healthcare}\label{sec:sec4C}
In healthcare, CFTel enables robotic-assisted surgery and assistive robotics, addressing the demand for remote interventions and patient support. CFTel facilitates real-time data processing to guide complex surgical procedures, incorporating advanced health technologies such as the Da Vinci Surgical System, and enhances surgical precision and dexterity~\cite{Dohler2024_1007, Kolb2025_MedRob}. Fog computing ensures low-latency control, critical for maintaining accuracy during telerobotic surgeries, achieving sub-millisecond response times~\cite{Hu2024_10595837}. Assistive robots are deployed in elderly care and rehabilitation for health monitoring and personalized mobility support in a time-sensitive manner.

CFTel’s distributed architecture enables scalable healthcare solutions and supports RaaS for broad deployment~\cite{Afrin2021_3061435, Kolb2025_MedRob}. However, challenges, including data privacy, cybersecurity, and regulatory compliance, present significant obstacles. Secure, privacy-preserving architectures are crucial to safeguard sensitive patient data while ensuring operational efficiency, prompting research into blockchain-based security and standardized protocols for healthcare robotics~\cite{Kolb2025_MedRob, Dohler2024_1007}.

\subsection{Disaster Management}\label{sec:sec4D}
CFTel is pivotal in disaster management, where swarm robotics performs search-and-rescue missions in hazardous environments. Cloud-fog computing enables real-time information sharing and coordination, while decentralized AI models at edge nodes facilitate autonomous decision-making~\cite{Zhang2024_104131}. Robots deployed in earthquake-affected areas navigate using EAI, coordinating via 5G to locate survivors. DTs monitor robot states remotely, allowing operators to optimize mission strategies through immersive interfaces~\cite{Tran2025_S2452414X24001869, Nguyen2024_10912837}. These capabilities enhance responsiveness in unpredictable disasters.

Challenges include achieving robust inter-robot communication under limited bandwidth and ensuring resilience in disrupted networks~\cite{Hu2024_10595837, Jin2024_3272696}. Future advancements in 6G networking, decentralized AI, and EAI are expected to improve swarm robotics efficiency, enabling more autonomous and scalable disaster response systems. CFTel’s integration of distributed intelligence and deterministic communication positions it as a critical technology for life-saving operations.

\section{Practical Challenges}\label{sec:sec5}
CFTel offers transformative potential for autonomous ICPS, but significant challenges hinder its widespread adoption. These obstacles span technical performance, security, and system interoperability, impacting the reliability, safety, and scalability of telerobotic applications. This section examines three critical challenges, details the implications, and proposes solutions to empower CFTel’s integration into various domains. Addressing these challenges is essential to realizing the promise of deterministic, autonomous, and service-oriented solutions through CFTel architecture.

\subsection{Latency and Real-Time Constraints}\label{sec:sec5A}
Achieving ultra-low latency and deterministic real-time communication is a formidable challenge for implementing CFTel, particularly in time-critical applications such as telerobotic surgery~\cite{Dohler2024_1007, Lopez2022_DT} and multi-robot coordination in disaster response~\cite{Afrin2021_3061435}. Although sub-millisecond response times are essential to ensure precision and safety, network congestion, jitter, and packet loss often introduce unacceptable delays~\cite{Hu2024_10595837}.

TSN provides deterministic Ethernet-based transmission, while 5G URLLC achieves approximately 1 ms latency with 99.999\% reliability~\cite{Yang2024}. Wi-Fi 6/7 further enhance throughput in dense environments~\cite{Ma2019_2907245}. However, these technologies struggle under dynamic conditions, such as fluctuating network loads in logistics hubs. AI-powered network orchestration can prioritize critical data streams, while latency-aware edge processing offloads computations to edge nodes, reducing transmission delays~\cite{Khan2020_1010, Zhang2024_104131}. Adaptive scheduling models enhance responsiveness by predicting congestion. Emerging 6G networks, AI-orchestrated network slicing, and Quantum Internet promise further improvements, but their development remains nascent, requiring further research.

\subsection{Cybersecurity}\label{sec:sec5B}
The CFTel architecture introduces cybersecurity risks, as data and control signals traverse multiple layers, exposing vulnerabilities to threats such as man-in-the-middle attacks, adversarial AI, and ransomware. Unauthorized access to telerobotic systems could disrupt operations and compromise sensitive data, with severe consequences in critical applications. For example, a breach in a surgical robot endangers patients~\cite{Dohler2024_1007}, while compromised logistics robots disrupt supply chains~\cite{Zhang2024_104131}.

Traditional security measures are vital, but they introduce latency and conflict with advanced requirements. Advanced solutions, such as blockchain-based authentication and zero-trust security models, offer robust protection but incur computational overhead. In multi-agent robotic systems, establishing trust among robots is complex, necessitating decentralized security models that verify authenticity without centralized reliance~\cite{Zhang2024_104131}. An AI-driven anomaly detection system can monitor behavior in real time, identifying unusual patterns indicative of attacks~\cite{Adil2025_3486659}. As quantum computing advances, quantum-resistant cryptographic techniques are needed to safeguard against future threats~\cite{Jin2025_04908}. Developing lightweight, low-latency security protocols is critical to balance protection and performance in CFTel for autonomous ICPS.

\subsection{Standardization}\label{sec:sec5C}
The absence of standardized frameworks and protocols poses a major barrier to the scalability and interoperability of CFTel. Proprietary architectures and communication protocols prevalent in industries lead to compatibility issues, complicating the integration of robotic systems from different vendors. This fragmentation hinders efficient data exchange and real-time collaboration, limiting the adoption of CFTel.

Interoperability challenges are particularly pronounced between open-source platforms and proprietary systems, which often use different data formats and communication protocols. For example, ROS relies on a publish-subscribe model, while proprietary systems may use custom APIs, leading to integration bottlenecks. In practice, protocols such as ROS~\cite{Xiao2024_111974}, OPC-UA~\cite{Ji2025_102927}, and Message Queuing Telemetry Transport (MQTT)~\cite{Luo2023_3228558} provide partial solutions but lack a unified architecture to integrate diverse robotic platforms, networking infrastructures, and AI-driven systems. A potential solution is integrating ROS 2 with OPC-UA to bridge robotic and industrial systems. Software-Defined Networking (SDN) and Network Function Virtualization (NFV) further address these challenges by providing a flexible network layer that abstracts the differences between heterogeneous systems, enabling seamless data exchange and control~\cite{Ma2019_2907245}. Standardization efforts by IEEE, ISO, and IEC are ongoing, but inconsistent adoption persists. Universal middleware solutions that abstract underlying complexities can ensure semantic interoperability.}

\section{Conclusion and Future Research Directions}\label{sec:sec6}
This state-of-the-art review has comprehensively explored CFTel and its pivotal role in advancing ICPS. By leveraging CFA, CFTel overcomes the limitations of conventional cloud-based telerobotics, delivering deterministic communication, computing, and control through the Cloud-Edge-Robotics architecture. The review details enabling technologies, including TSN, 5G URLLC, Wi-Fi 6/7, EI, EAI, and DTs.

CFTel impact spans diverse applications, including adaptive automation in smart manufacturing and precision farming in agriculture, to efficient robotic fleets in logistics and warehousing, precise interventions in healthcare, and resilient operations in disaster management. Despite these advancements, challenges such as latency, cybersecurity, and standardization persist, necessitating innovative solutions to ensure reliability and scalability. This paper serves as a foundational resource for researchers and industry professionals, synthesizing recent advancements and providing a roadmap for future innovation in telerobotics, thereby contributing to the evolution of autonomous and intelligent industrial automation.

Further research is required to fully leverage CFTel's potential, address existing constraints, and utilize emerging technological advancements to facilitate next-generation autonomous and secure industrial systems. Key directions include AI-driven time-sensitive operations, deterministic networking, sustainable distributed computing paradigms, advancements in EI and EAI, cybersecurity and privacy solutions, standardization and interoperability, and application-specific challenges.

\bibliographystyle{ieeetr}
\bibliography{Bibliography}

\end{document}